\documentclass[aps,pre,twocolumn,showpacs]{revtex4}
\usepackage{graphicx}
\usepackage{wasysym}

\newcommand{\be}{\begin{equation}}
\newcommand{\ee}{\end{equation}}
\newcommand{\bea}{\begin{eqnarray}}
\newcommand{\eea}{\end{eqnarray}} 
\begin{document}

\title{Mean-field limit of systems with multiplicative noise}
\author{Miguel A. Mu\~noz}
\affiliation{Departamento de Electromagnetismo y F{\'\i}sica de la Materia and
  Instituto Carlos I de F{\'\i}sica Te{\'o}rica y Computacional
\\ Facultad de
  Ciencias, Universidad de Granada, 18071 Granada, Spain}
\author{Francesca Colaiori}
\affiliation{Dipartimento di Fisica, Universit\`a di
Roma ``La Sapienza'' and Center for Statistical Mechanics and Complexity,
INFM Unit\`a Roma 1, P.le A. Moro 2, 00185 Roma, Italy}
\author{Claudio Castellano}
\affiliation{Dipartimento di Fisica, Universit\`a di
Roma ``La Sapienza'' and Center for Statistical Mechanics and Complexity,
INFM Unit\`a Roma 1, P.le A. Moro 2, 00185 Roma, Italy}
\date{\today}

\begin{abstract}
  A detailed study of the mean-field solution of Langevin equations
  with multiplicative noise is presented. Three different regimes
  depending on noise-intensity (weak, intermediate, and strong-noise)
  are identified by performing a self-consistent calculation on a
  fully connected lattice. The most interesting, strong-noise, regime
  is shown to be intrinsically unstable with respect to the inclusion
  of fluctuations, as a Ginzburg criterion shows. On the other hand,
  the self-consistent approach is shown to be valid only in the
  thermodynamic limit, while for finite systems the critical behavior
  is found to be different. In this last case, the self-consistent
  field itself is broadly distributed rather than taking a well
  defined mean value; its fluctuations, described by an effective
  zero-dimensional multiplicative noise equation, govern the critical
  properties. These findings are obtained analytically for a fully
  connected graph, and verified numerically both on fully connected
  graphs and on random regular networks. The results presented here
  shed some doubt on what is the validity and meaning of a standard
  mean-field approach in systems with multiplicative noise in finite
  dimensions, where each site does not see an infinite number of
  neighbors, but a finite one. The implications of all this on the
  existence of a finite upper critical dimension for multiplicative
  noise and Kardar-Parisi-Zhang problems are briefly discussed.
\end{abstract}

\pacs{02.50.Ey,05.50.-a,64.60.-i}

\maketitle

\section{Introduction}

 Problems susceptible to be mathematically represented by stochastic
 (Langevin) equations including a multiplicative noise abound not only
 in physics, but also in biology, ecology, economy, or social
 sciences. In a broad sense a Langevin equation is said to be
 multiplicative if the noise amplitude depends on the state variable/s
 itself/themselves \cite{vK}. In this sense, problems exhibiting
 absorbing states, i.e. fluctuation-less states in which the system
 can be trapped, are described by equations whose noise amplitude is
 proportional to the square-root of the (space and time dependent)
 activity density, vanishing at the absorbing state \cite{AS}. Systems
 within this class are countless: propagating epidemics, autocatalytic
 reactions, reaction-diffusion problems, self-organized criticality,
 pinning of flux lines in superconductors, etc. \cite{AS}.

 In a more restrictive sense, the one we will use from now on, it is
 customary to restrict the term {\it multiplicative noise} (MN) to
 noise amplitudes {\it linear} in the activity density
 \cite{Graham,Sancho,Redner,Sornette}. Such equations appear
 ubiquitously in economics, optics, population dynamics, study of
 instabilities, etc.  In all these cases, multiplicative noise terms
 appear rather straightforwardly when constructing (more or less
 rigorously) stochastic representative equations.  In their spatially
 extended version, Langevin equations including a MN (see
 Eq.~(\ref{mn}) below) were first proposed, to the best of our
 knowledge, in the context of synchronization of coupled map lattices
\cite{PK}, and soon after studied in Refs.
\cite{MN,Walter,vdB,Marsili,Birner}. Such equations describe different
situations; in particular, there has been a recent interest in their
application to non-equilibrium wetting \cite{Lisboa} and also to
synchronization problems in extended systems \cite{Synchro}; see
Ref. \cite{MNreview} for a recent review. In all these cases, there is
a phase transition between an active phase in which the system has a
stationary non-vanishing activity and another one in which the density
field falls continuously toward zero without ever reaching it in
finite time. The analogies and differences between this family of
models and the one with a square-root type of noise, in which the
absorbing state is reachable within a finite time, have been discussed
in \cite{nature,Janssen}.

 It is remarkable that by using the so-called Cole-Hopf
 transformation, the spatially extended MN Langevin equation can be
 mapped into a non-equilibrium interface, as represented by the
 Kardar-Parisi-Zhang (KPZ) equation \cite{KPZ,HZ} in the presence of a
 bounding wall \cite{MNreview}. In this language, the active phase
 describes interfaces pinned by the wall while the absorbing one
 corresponds to depinned KPZ-like interfaces moving away from the
 wall. 

 In recent years many aspects of systems with MN have been elucidated.
 For example, a renormalization group approach has been constructed,
 scaling relations derived, and the large-$N$ limit (where $N$ is the
 number of components) studied \cite{MN}. It is well established that
 above two-dimensions there are two different regimes; a weak-noise
 and a strong-noise one, in full analogy with the known phenomenology
 of KPZ interfaces \cite{HZ}.  There is however a crucial point which
 remains to be fully understood: the mean-field behavior of such
 systems. Another interesting mapping is that in the absence of the
 non-linear term, the MN equation corresponds to the equation
 governing the evolution of directed polymers in random media (See
 \cite{HZ} and references therein, as well as \cite{MNreview}).

 In Langevin equations with additive noise, as for instance the Model
 A describing the universality class of Ising like transitions
 \cite{HH}, the {\it mean-field} approximation can be obtained in a number
 of equivalent ways, all of them leading to the same results with
 different degrees of sophistication. For example, mean-field critical
 exponents can be obtained (i) by discarding the noise and solving the
 remaining deterministic equation, (ii) as the most probable solution in a
 path integral formulation, (iii) self-consistently by assuming that each
 site sees the average of the remaining, (iv) by naive power counting in
 the corresponding action, (v) as the lowest order in a perturbative loop
 expansion, etc.

 Contrarily, in systems with MN the meaning of the mean-field solution
 is much less clear. Indeed, early studies \cite{Walter} led to
 conflicting results depending on the considered approximation method.
 For example, by removing the noise, much of the physics is lost and
 trivial results (identical to those for additive noise) are obtained.
 Also, if all spatial dependence is eliminated (by removing the
 Laplacian term in the MN equation), one is left with a solvable
 zero-dimensional equation which does not reproduce faithfully the
 rich mean-field phenomenology. Therefore, contrarily to more standard
 problems, both space-dependence and noise have to be retained in
 order to construct a sound mean-field solution. 

  If the bounding wall is eliminated then one is left with the {\it
  directed polymer in random media} equation, for which many results
  are available. In particular, Derrida and collaborators worked out a
  solution on the Cayley tree, mean-field results, $1/d$ expansions,
  and solutions on hierarchical lattices exits \cite{Derrida}. Also,
  M\'ezard and Parisi derived a variational approach in replica space
  \cite{MP}, and also Fisher and collaborators reached also
  interesting results on these issues \cite{Fisher}. For a more
  detailed review on results for directed-polymers see \cite{HZ} and
  references therein.

  On the other hand, restoring the bounding wall (which is the case we
  are interested in) Birner et al. \cite{Birner} (see also
  \cite{Walter,vdB,Marsili}) performed a self-consistent calculation
  and reported on the existence of two different mean-field behaviors:
  a weak-noise regime (in this particular case, the noise can be
  completely disregarded) and a strong-noise one exhibiting
  non-universal exponents depending on the noise amplitude. This is in
  agreement with the field theoretic expectation of two different
  behaviors for MN-like and KPZ-like equations in high dimensional
  systems (where mean-field results are expected to be valid)
  \cite{HZ,MNreview}. These results are difficult to compare with the
  abovementioned ones obtained for directed polymers in random media,
  which correspond to the deppined phase of the full problem, while
  the self-consistent approach is intrinsically devised for the pinned
  phase.
 
  Let us also underline that a full understanding in terms of path
  integrals and extremal paths is still missing despite some efforts
  in this direction \cite{Kharchenko,Fogedby}.

  Together with the determination of the right mean-field theory,
  another relevant and highly debated issue is to establish the upper
  critical dimension $d_c$, above which mean-field results hold.  As
  some exponents, as the dynamical one $z$ have been claimed
  \cite{MN,MNreview} to coincide for MN and KPZ, both problems are
  expected to have the same upper critical dimension. While there is
  consensus that the mean-field weak-noise regime should be valid
  above $d_c=2$ (coinciding with the critical dimension for weak-noise
  Edwards-Wilkinson interfaces \cite{HZ}) there exist highly
  conflicting results (some pointing to $d_c=4$ \cite{KPZdc4} and some
  supporting $d_c=\infty$ \cite{KPZdcinf}) for the strong-noise one
  regime.

 In any generic problem, for sufficiently high dimensions every site
 in a spatially extended system ``sees'' the average of its neighbors
 (assuming the space has been discretized) which can be correctly
 approximated by its mean value (the distribution of the average
 values is well described by the standard central limit theorem above
 the upper critical dimension) defining in this way a sound mean-field
 solution. In the case of MN, as we will show, the situation is
 somehow anomalous: in the strong-noise regime the mean-field solution
 itself breaks down in the neighborhood of the critical point once
 fluctuations are taken into account (as a version of the Ginzburg
 criterion shows). This stems from the primary role played by
 fluctuations and rare events in multiplicative processes
 \cite{Redner,Sornette}, and makes one wonder whether a mean-field
 solution can be valid at all in any finite dimension.
  
 In this paper we revisit the self-consistent mean-field solution of
 Langevin equations with MN and report on a new previously overlooked
 regime. After that, we discuss under which circumstances such a
 solution breaks down (Ginzburg criterion \cite{Birgenau,LB}).  Also, we
 present a fluctuating-solution aimed to extend the self-consistent
 one by allowing for fluctuations of the average field. Finally, we
 present some speculations on the issue of the existence of an upper
 critical dimension for this type of systems, i.e. if there is a
 finite space dimensionality above which the non-fluctuating solution
 holds or not.

 The paper is organized as follows. In Section~\ref{SecNF}, we outline
 a self-consistent solution of the MN equation defined in an infinite
 fully connected lattice. Different regimes are found, corresponding
 to weak-noise, intermediate-noise, and strong-noise respectively. By
 constructing a Ginzburg criterion we will show how the
 strong-noise solution is intrinsically unstable as soon as
 fluctuations are taken into account. In Section~\ref{SecFS} we study
 the solution in finite lattices (fluctuating solution) and show by
 means of numerical simulations how the mean-field solution breaks
 down in the vicinity of the critical point. Section~\ref{SecRRG} is
 devoted to the study of multiplicative noise on random regular
 graphs. This is done in order to ascertain the type of behavior in
 the thermodynamic limit when the connectivity of each site remains
 finite and, therefore, in order to obtain some insight on the
 behavior of the MN equation in large but finite space dimensions. The
 final section contains a summary of the findings along with some
 concluding remarks. Finally, a general scaling theory is presented in
 an appendix.

\section{Non Fluctuating Solution}
\label{SecNF}

We summarize and extend the mean-field solution obtained by Birner et
al. \cite{Birner} for the MN equation as described in
\cite{MN,MNreview}:
\be {\dot \phi}_i = a
\phi_i -b \phi_i^{p+1} + D \nabla^2 \phi_i + \phi_i \eta_i \sigma
\label{mn}
\ee where $\eta$ is a Gaussian delta correlated noise, $a$, $b$, $D$,
and $\sigma$ are constants, and $\phi$ is the local order-parameter
field, describing the physical density under study. In order to study
its mean-field solution, we consider the case of global coupling,
i.e. define the system on a fully connected (or complete) graph
\be \nabla^2 \phi_i = {1 \over N-1} \sum_{j \neq
i} (\phi_j - \phi_i) \equiv M_i -\phi_i.
\label{lapla}
\ee 
The associated Fokker-Planck equation \cite{vK} for Eq.~(\ref{mn}) in
the Ito sense can then be solved in the stationary
case~\cite{Birner,Walter,Marsili}, assuming $M_i=M$, \be R(\phi;M) =
\phi^{-2-2(D-a)/\sigma^2} \exp\left[- {2b \phi^p \over p \sigma^2} -{2
M D \over \phi \sigma^2 } \right].  \label{R} \ee This is a power-law
distribution function with two cutoffs: an upper one coming from the
non-linear saturation term and a lower one generated by $M$ which
acts as a constant external field.

In order to proceed further $M$ is taken equal to its average over
realizations, $m$, which is determined self-consistently by
imposing~\cite{vdB,Walter,Birner}
\be
m = \langle \phi \rangle = {\int d\phi \phi R(\phi;m) \over \int d\phi
R(\phi;m)}.
\label{selfcon}
\ee 

We will denote in the following this solution, valid for $N=\infty$,
as the Non Fluctuating (NF) solution.  Equation~(\ref{selfcon}) has
always the trivial solution $m=0$, stable for $a<a_c$ with $a_c=0$. At
the critical point, $a_c$, a stable solution with $m>0$ appears.
Introducing the distance from the critical point $\epsilon=
a/(\sigma^2/2)$ and the reduced variable $s=2D/\sigma^2$, we can
rewrite the probability distribution as
\be R(\phi|m) = \phi^{-\alpha}
\exp\left[- {b s \phi^p \over p D} - {m s \over \phi } \right]
~~~~~~~~~~~\alpha = 2+s -\epsilon.  \ee 
Introducing the notation \be
I_{s,k} = \int_{0}^\infty d\phi
\, \phi^k R_s(\phi|m) \ee the first moment $m=I_{s,1}/I_{s,0}$ is
easily computed, yielding the relation between $m$ and $\epsilon$ in
the active region \be m^{\min[s-\epsilon,p]} \sim \epsilon \ee so that
$m \sim \epsilon^{\beta_1}$ with $\beta_1=\max[1/s,1/p]$.  In this
way, two different regimes were found by Birner et al.: one is
universal (weak-noise) and the other one is not (strong-noise) with
the order-parameter exponent changing continuously with $s$.  These
two regimes are the analogue of the well-known weak-coupling
(Gaussian) and strong-coupling (non Gaussian) regimes of KPZ dynamics
\cite{HZ}.

Let us now go beyond the results in \cite{Birner} by computing higher
moments $m_k = \langle \phi^k \rangle = I_{s,k}/I_{s,0}$.  The
normalization $I_{s,0}$ scales as $m^{-1-s+\epsilon}$, and to leading
order \be I_{s,k}=I_{s-k+1,1}
\simeq B_k m^{-s+k-1+\epsilon}+C_k + D_k m^{p-s+k-1+\epsilon}.  \ee
While the third term is always sub-leading, it depends on $k$ which of
the other two dominates.  For $s>k-1+\epsilon$ the first one
is dominant, so that \be m_k = \langle \phi^k \rangle \sim {
m^{k-1-s+\epsilon}\over m^{-1-s+\epsilon}} \sim { m^k}
\label{m_k}
\ee
and for $s<k-1+\epsilon$ the second is the leading one, so that
\be
m_k = \langle \phi^k \rangle \sim {1 \over m^{-1-s+\epsilon}} 
\sim m^{1+s-\epsilon}
\label{m_k2}
\ee 
Hence, the regime with non universal first moment is quite rich.  For
$1/p<1/s<1$ the first $k$ moments exhibit standard scaling, as long as
$k<s+1-\epsilon$, while all others scale with the same exponent
$1/s+1$. We call this regime {\it intermediate-noise}; it does not
have a clear analogue in free KPZ-like interfaces.

\begin{figure}
\includegraphics[angle=0,width=8cm,clip]{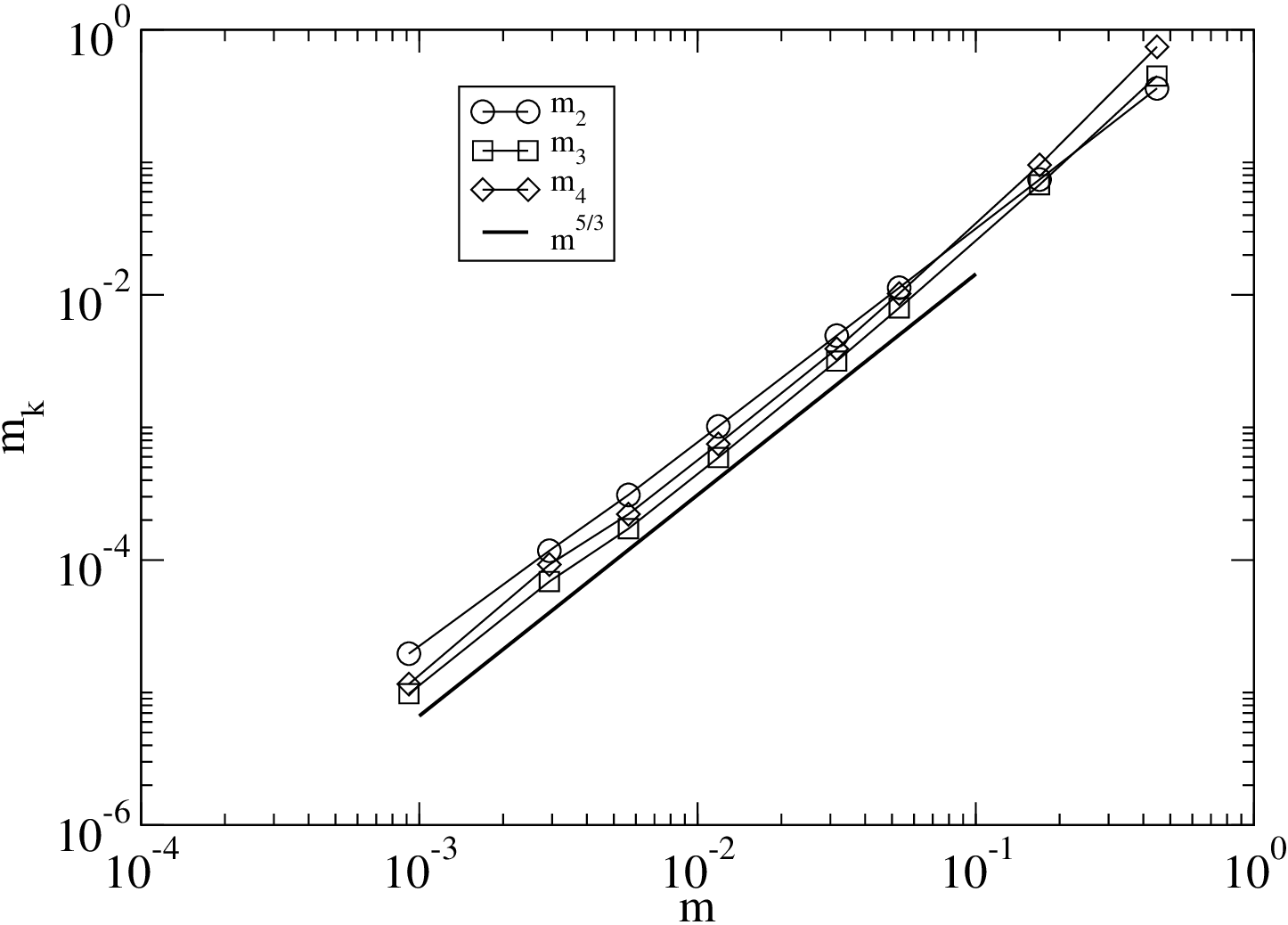}
\caption{Plot of the second, third and fourth moment as a function of
the first one in the NF strong-noise regime, for a system with $p=2$,
$N=5000$ and $s=2/3$. All moments scale in the same way.}
\label{Fig1}
\end{figure}

When $1/s >1 $ full multi-scaling occurs: all moments, except the
first, scale with the same exponent, $1/s+1$.  This is the {\it
strong-noise} regime; numerical evidence of this multi-scaling is
provided in Fig.~\ref{Fig1}, and is similar to the known phenomenology
of single-site (zero-dimensional) MN equations \cite{Graham,MN}.

>From naive power counting (usually expected to reproduce mean-field
exponents) performed on the MN equation, the naive time scale $T$
should scale as $T^{-1} \sim a \propto \epsilon$ and, therefore, we
expect the critical time-decay exponent of the order parameter,
$\theta_1$ (or simply $\theta$) defined by $(\langle
\phi(t,\epsilon=0) \rangle
\sim t^{-\theta_1}$) to coincide with $\beta_1$ in mean-field. The
same property remains valid for higher moments, so the multi-scaling
property of the static exponent is translated into multi-scaling of
the decay exponents \cite{footnote}.

Let us now construct a Ginzburg criterion within the fully
connected lattice, to see under which circumstances the previous
approximation ceases to be sound. To do so we just need to compute the
ratio $r \sim m_2 \over m_1^2$  \cite{Birgenau,LB}. 
Whenever $r$ diverges fluctuations are expected to play a significant
role, breaking down the self-consistent solution which ignores them,
as soon as they are taken into consideration. From the previously
reported scaling for the strong-noise solution one has $r\sim
\epsilon^{1-1/s}$, which diverges at the critical point if $1/s > 1$,
{\it i. e.} in any case cite \cite{GGC}.

Therefore, the previously reported strong-noise mean-field solution is
fully valid only in the strict $N=\infty$ limit, when fluctuations can
be safely discarded owing to the law of large numbers. In general, it
is expected that such a strong-noise solution will break down as soon
as fluctuations are somehow taken into account. If, for example, the
MN equation is defined on the top of a d-dimensional lattice (every
site sees a finite number of others) the NF strong-noise solution is
not expected a priori to be valid. We will discuss this aspect more in
detail in the forthcoming sections. On the other hand, for the weak-
and the intermediate-noise regimes, we have Gaussian scaling of the
lowest moments, and therefore $r$ converges to a constant at the
critical point.

\section{Fluctuating Solution}
\label{SecFS}

In the previously reported approach the crucial step is the
replacement of $M$ (the mean value seen by any arbitrary single site)
by a fixed non-fluctuating $m$. As said before, owing to the law of
large numbers this is exact if $N = \infty$ whatever the probability
distribution of the neighboring sites, but one can wonder whether this
replacement is acceptable if a finite system size $N$ is
considered. In other words: what is the critical behavior for large
but finite values of $N$?  This is a perfectly sensible question,
since for MN a sharp phase transition is well defined for any value of
$N$, even for $N=1$
\cite{Graham}.

In this section we relax the condition of a fixed value for $M$, by
allowing the spatial average value of the $\phi$ field in the
stationary state to fluctuate in time, sampling some probability
distribution $Q(M)$ to be determined self-consistently.

Assuming that $M$ changes in a time scale much larger than the
characteristic time scale for $\phi$, we can still solve the
Fokker-Planck for a fixed value of $M$, and therefore Equation~(\ref{R})
still holds, but now it has to be interpreted as a conditional
probability $R(\phi \mid M)$.  The full distribution $P(\phi)$ is
given by the convolution: \be P(\phi) = \int_0^{\infty} dM R(\phi|M)
Q(M).  \ee 

In this case, the self-consistent Equation~(\ref{selfcon}) has to be
replaced by a self-consistent functional equation for $Q(M)$.  

\be
\int dM ~ M~ Q(M) = \frac{ \int d \phi~ \phi~  \int dM Q(M) R(\phi \mid
M)} { \int d \phi~  \int dM Q(M) R(\phi \mid M)}. \label{functional}
\ee
By solving this functional equation one could obtain the full solution
as in the NF case, within the slow changing $M$ approximation.

Instead of solving numerically such a self-consistent equation we now
try to write down an evolution equation for $M(t)$, from which $Q(M)$
follows. For that, we sum Eq.~(\ref{mn}) over $i$ and divide it by
$N$, giving
  \be {\dot M} = a M - b {1 \over N} \sum_i
\phi_i^{p+1} +
\sigma {1 \over N} \sum_i \phi_i \eta_i.
\label{eqM}
\ee
This turns out to be an effective MN equation for the field $M$ in
zero dimensions. Let us justify this statement.  The second term on
the r. h. s. is perfectly analogous to the nonlinear term in
Eq.~(\ref{mn}): it just introduces an upper cutoff $M_u$ in the
distribution of the field $M$.  The last term on the r. h. s. is less
trivial: it can be rewritten as $\sigma {1 \over N} \sum_i {\phi_i
\over M} \eta_i M$, hence in the form of a multiplicative noise $\eta'
M$.  The average value of $\eta'$ is clearly zero, since $\eta'$ has
random sign.  Its second moment, that we denote as $\sigma^2(N)$, is
\be
\sigma^2(N) =  { \sigma^2 \over N^2} \langle 
(\sum_i {\phi_i \over M} \eta_i)^2 \rangle.
\ee
The variables $\eta_i$ and $\phi_i$ are uncorrelated.  The $\eta_i$
are normally distributed.  In practice, to compute $\sigma^2(N)$ we
must evaluate the sum of $N$ variables $\phi_i/M$, distributed
according to $R(\phi/M|M)$ with random signs. The distribution
$R(\phi/M|M)$ is a power-law with exponent $\alpha=2+s-\epsilon$,
lower cutoff in 1 and upper cutoff in $1/M$.

We now discuss the properties of this solution depending on
whether we are working in the strong-noise regime or not.

\subsection{Weak- and Intermediate-Noise}
Let us consider first what happens when $1/s<1$, i.e. in the weak and
the intermediate noise regimes. In these cases, the exponent $\alpha$
of the distribution $R$ is larger than $3$, for $a$ sufficiently
small.  Adding $N$ such variables with random signs is equivalent to
adding Gaussian variables \cite{Levy}. Then \be \left(\sum_i {\phi_i
\over M} \eta_i \right)^2 \sim N, \ee so that \be \sigma^2(N) \sim
\sigma^2 /N.  \ee We can conclude that the dynamics of $M$ is governed
by an effective equation with MN in zero dimensions with a noise
$\eta'$ with renormalized variance $\sigma^2/N$.  It is clear that
$\eta'$ is correlated in time, but its finite correlation-time can be
eliminated by suitably rescaling the time variable.  The distribution
for $M$ is then given by the solution of the zero-dimensional
MN~\cite{Graham} \be Q(M) = M^{-2[1-a/\sigma^2(N)]} \exp(-M/M_u)^2.
\label{Qdist}
\ee 
As expected, this distribution function is a solution of the previously written
functional self-consistent Equation~(\ref{functional}).

For finite $N$ the system undergoes an absorbing phase transition for
a finite value of the control parameter
\be
a_c(N) = {\sigma(N)^2 \over 2} \sim {\sigma^2 \over 2 N}.
\label{acGaussian}
\ee
The exponents for the transition with finite $N$ are given by the
values for zero-dimensional MN transition~\cite{Graham,MN}: $\beta_1=1$,
and $\theta=1/2$.

Notice that this transition is qualitatively different from that
occurring in the thermodynamic limit.  Here the transition takes place
because the distribution of $M$ becomes non-normalizable due to the
divergence for $M \to 0$.  In the thermodynamic limit, instead, the
distribution $Q(M)$ remains normalizable (Gaussian) at the transition
and criticality comes from the peak position $M_u$ moving toward zero
(see the appendix, where a coherent general scaling picture is
presented).

These two regimes are therefore distinguished by the presence of a
(Gaussian) peak for finite $M$ (the NF limit) or a broad distribution
with a power-law divergence for $M \to 0$.  The crossover occurs where
the power-law exponent in Eq.~(\ref{Qdist}) changes sign, i.e. for \be
{\tilde a}(N) = \sigma^2(N) = 2 a_c(N).  \ee The behavior for fixed
$N$ can then be summarized as follows.  For $a \gg {\tilde a}(N)$ the
system exhibits the critical behavior of the weak noise $N = \infty$
solution.  For $0 < a -a_c(N) \ll {a_c}(N)$ it behaves as it was
zero-dimensional.  In the latter case, the crossover can also be
observed with fixed $a$ by looking at the temporal evolution of the
first moment $m$: at short times the NF solution is followed,
crossing-over later to the asymptotic zero-dimensional scaling.

\begin{figure}
\includegraphics[angle=0,width=8cm,clip]{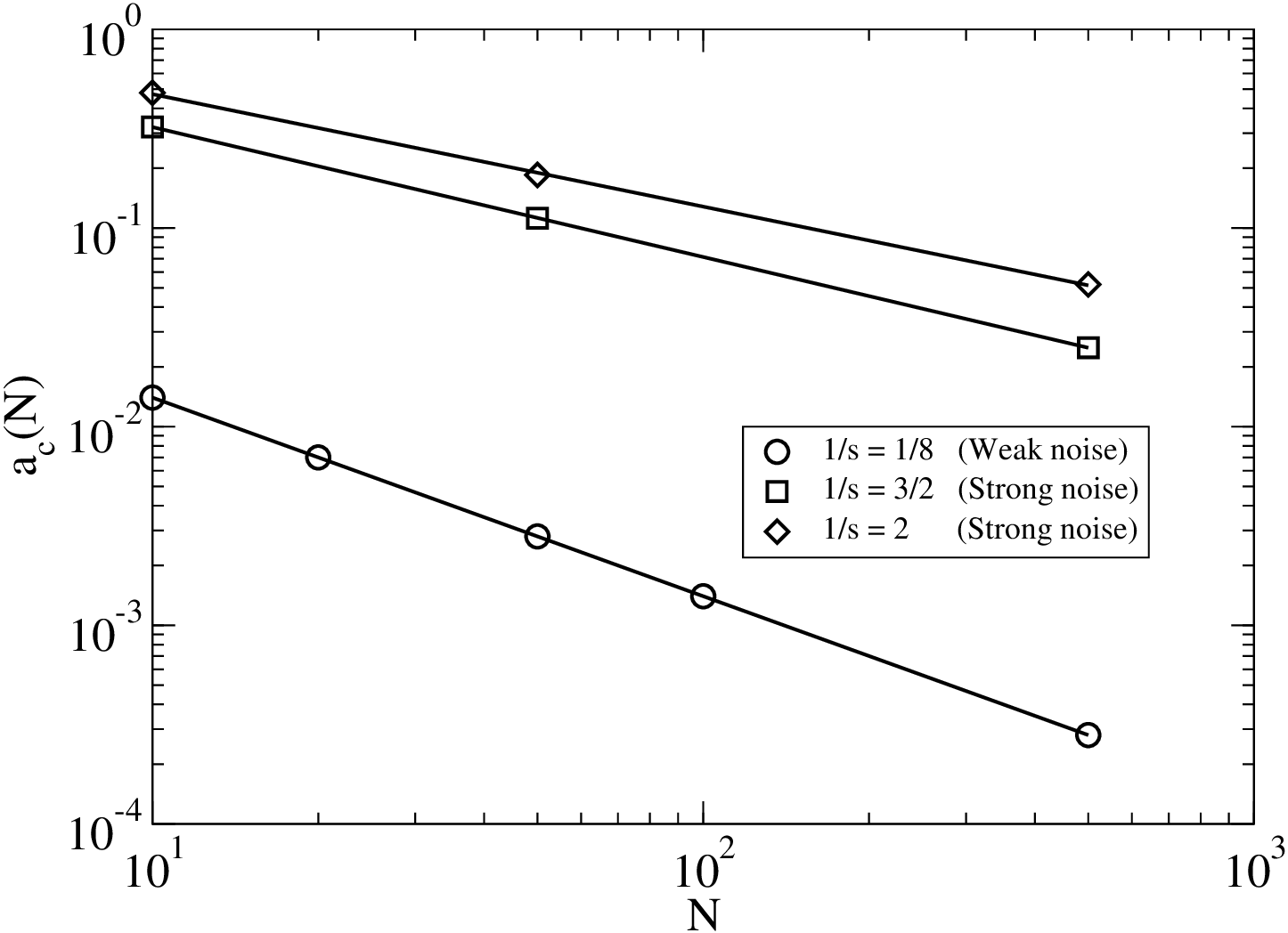}
\caption{Plot of $a_c(N)$ versus $N$ for three values of $1/s$.
Thick lines are power-laws fits with exponents $-1$ for $1/s=1/8$,
$-0.65$ for $1/s=3/2$ and $-0.57$ for $1/s=2$ (to be compared with the
output of~Equations (\ref{acGaussian}) and (\ref{a_cSC})).}
\label{a_cvsN}
\end{figure}

We have checked the correctness of this scenario by means of numerical
simulations of Eq.~(\ref{mn}) with $p=2$.  From the numerical point of
view, the first problem is the determination of the critical point
$a_c(N)$.  The criterion we have chosen is based on the way the first
moment $m(t)$ decays in time.  $a_c(N)$ is the value separating a
concave behavior (for $a>a_c(N)$) from a convex one (for
$a<a_c(N)$). In this way, we obtain the values plotted in
Fig.~\ref{a_cvsN}. The behavior found in the weak-noise regime is in
perfect agreement with the predicted $1/N$ behavior.
 
\begin{figure}
\includegraphics[angle=0,width=8cm,clip]{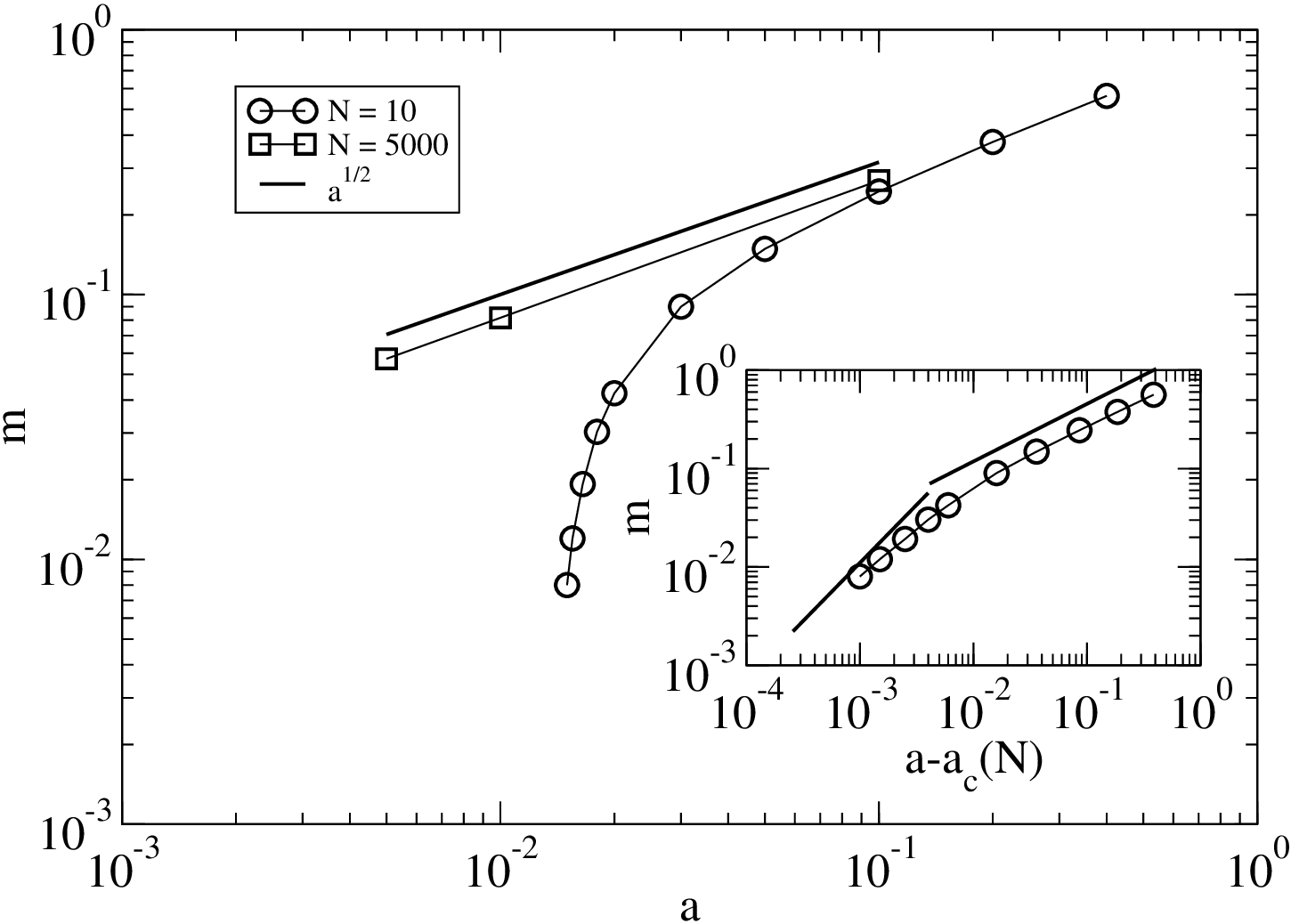}
\caption{Main: plot of $m$ vs $a$ for $1/s=1/8$ (weak-noise).
Inset: the same data of the main part with $N=10$, plotted vs
$a-a_c(N)$ and showing the NF behavior ($m \sim (a-a_c(N))^{1/2})$ and 
the zero-dimensional one ($m \sim (a-a_c(N))$). Solid lines are guides
to the eye.}
\label{m_WC}
\end{figure}

With $p=2$, the temporal behavior in the weak-noise case is the same
($\theta=1/2$) both in the thermodynamical limit and for the effective
zero-dimensional behavior valid at finite $N$, not allowing to
distinguish between the two regimes.  In Fig.~\ref{m_WC} we analyze
the behavior of the stationary value of the first moment $m$.  In the
main part of the figure we observe that, for $N=5000$, the system
follows very accurately (for the values of $a$ considered) the decay
with $\beta_1=1/2$, expected in the thermodynamic limit.  For $N=10$,
instead, a singularity is observed for finite $a$.  If we plot $m$ as
a function of $a-a_c(N)$ (Fig.~\ref{m_WC}, inset) both behaviors can
be observed: for $a-a_c(N) \gg a_c(N) \approx 0.014$ the scaling is of
the NF type ($\beta_1 = 1/2$).  For $a-a_c(N) \ll a_c(N)$ we observe
the zero-dimensional exponent $\beta_1=1$.

\begin{figure}
\includegraphics[angle=0,width=8cm,clip]{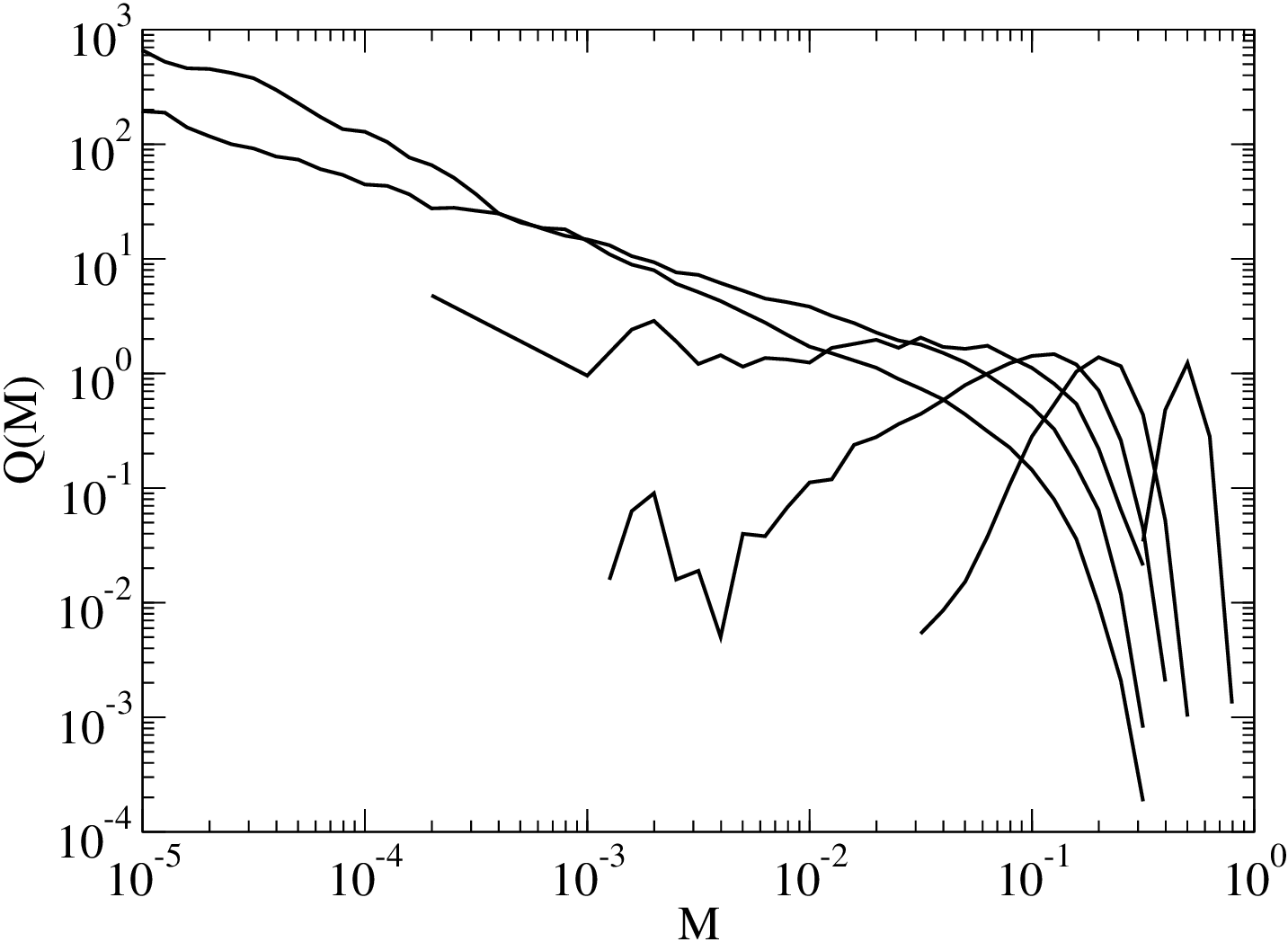}
\caption{Plot of $Q(M)$ vs $a$ for $1/s=1/8$ (weak noise), displaying 
the crossover between a Gaussian around $M_u>0$ and a power-law
divergence in $0$.  The value of the critical point is $a_c(N=10)
\approx 0.014$. The values of $a$ are, from right to left: $0.4, ~0.1,
~0.05, ~0.03, ~0.02, ~0.155$.}
\label{Q_WC}
\end{figure}

A direct validation of Eq.~(\ref{Qdist}) is provided by
Fig.~\ref{Q_WC}, where the distribution of the self-consistent field
$M$ exhibits a peak in $M_u(a)$ for $a-a_c(N) \gg a_c(N)$ while a
power-law divergence at zero develops for $a-a_c(N) \ll a_c(N)$.

\subsection{Strong-Noise}

Let us consider now the strong-noise case $1/s>1$.  The general
picture is similar to the one previously described, with the main
difference that the exponent $\alpha$ in the distribution
$R(\phi/M|M)$ is now smaller than $3$.  From the theory of Levy-stable
distributions~\cite{Levy} we know that the sum of $N$ variables
distributed as a power-law with exponent $1+\mu$ with $1<\mu<2$ and an
upper cutoff $1/M$, scales as $N^{1/\mu}$ for $N \ll M^{-\mu}$ while
it behaves in a Gaussian way, $N^{1/2}$, for $N \gg M^{-\mu}$.  This
implies that, for fixed $N$, $\eta'$ is a power-law distributed noise
with exponent $1+\mu$ for $M \ll N^{-1/\mu}$, and a Gaussian noise
with $\sigma^2(N)
\sim \sigma^2 /N$ for $M \gg N^{-1/\mu}$.  In the present case
$\mu=\alpha-1=1+s-2a/\sigma^2$.

Eq.~(\ref{eqM}) describes now a zero-dimensional MN with a rather
exotic noise, whose distribution depends on $M$. We have no clear
theoretical understanding of such a model.  In principle it could give
rise to completely new and non-trivial critical features. However, as
shown numerically below, it turns out to behave asymptotically as the
zero-dimensional case with standard noise.  The only change is in the
position of the critical point $a_c(N)$.  Accordingly, we assume,
rather crudely, that the power-law noise does not change the behavior
of the zero-dimensional MN, except for the form of $\sigma^2(N)$
\be \sigma^2(N) \sim \sigma^2
N^{2(1/(\alpha-1)-1)} = \sigma^2 N^{2[1/(1+s-2a/\sigma^2)-1]}.
\ee 
The distribution of $M$ is then given again by Eq.~(\ref{Qdist}), with
the additional complication that $\sigma^2(N)$ depends on $a$.  The
critical point is determined by the implicit condition \be a_c(N) =
{\sigma^2(N) \over 2} \sim \sigma^2 N^{2\left[\displaystyle {1 \over
(1+s-2 a_c(N)/\sigma^2)}-1\right]}.
\label{a_cSC}
\ee
If, as a first approximation, we neglect the dependence of $\alpha$ on
$a$, i.e. we take $\alpha=2+s$, we obtain $a_c(N) \sim N^{-2s/(1+s)}$,
not far from the numerical results of Fig.~\ref{a_cvsN}.

\begin{figure}
\includegraphics[angle=0,width=8cm,clip]{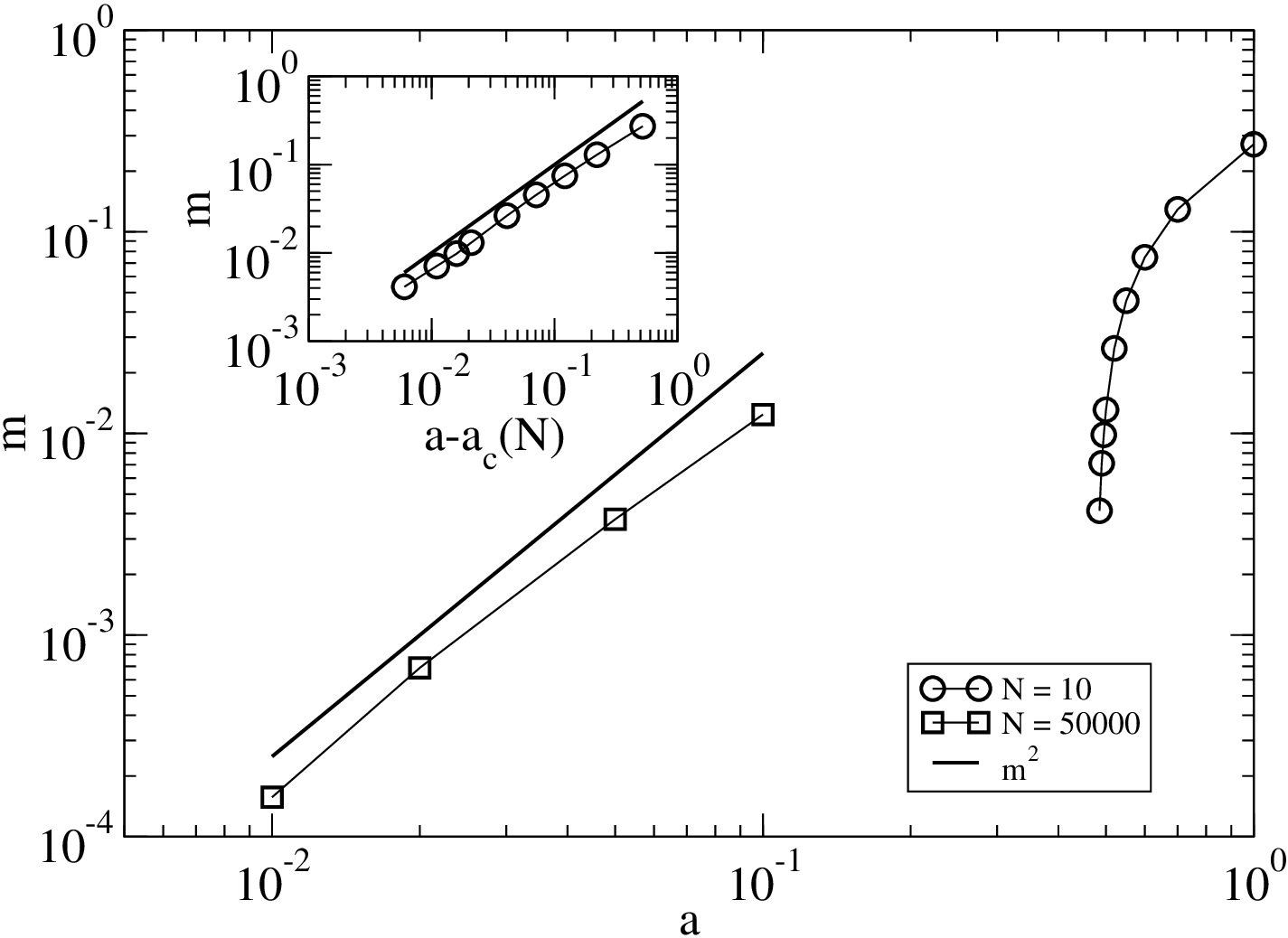}
\caption{Main: plot of $m$ vs $a$ for $1/s=2$ (strong-noise).
Inset: the same data of the main part with $N=10$, plotted vs
$a-a_c(N)$ and in agreement with the zero-dimensional linear behavior.
Solid lines are guides to the eye.}
\label{m_SC}
\end{figure}

Again for $a>{\tilde a}(N) = 2 a_c(N)$ the exponent of the
distribution $Q(M)$ becomes negative and there is a crossover to the
NF limit.  In Fig.~\ref{m_SC} we plot $m$ versus $a$ in the
strong-noise regime. Again the value $\beta_1=1/s$, valid in the
thermodynamic limit, is observed for large $N$, as the critical point
and the crossover are very close to $0$.  Near the transition the
exponent is the zero-dimensional one, $\beta_1=1$, indicating that the
non trivial noise does not modify the zero-dimensional critical
behavior.
 
\begin{figure}
\includegraphics[angle=0,width=8cm,clip]{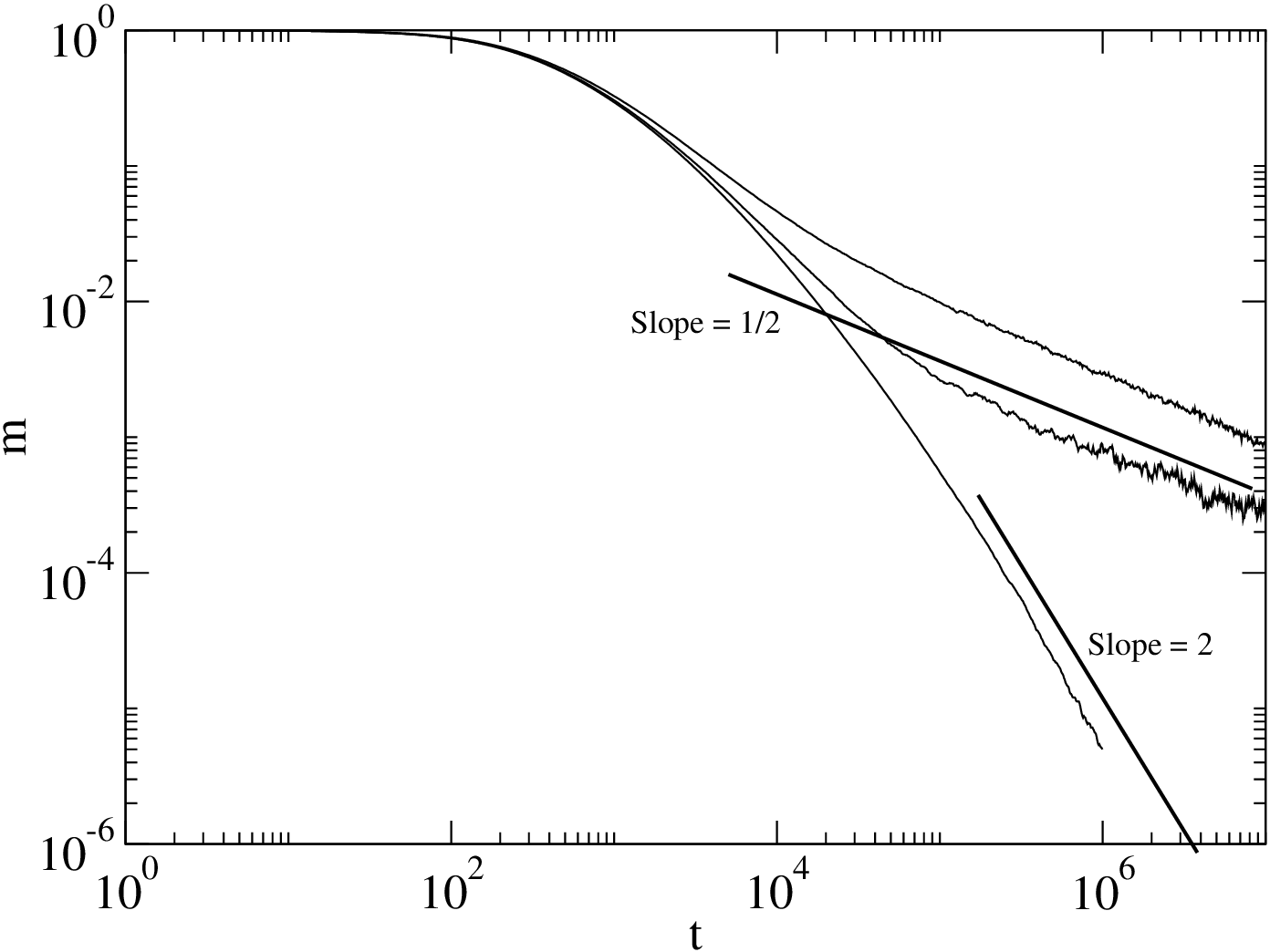}
\caption{Plot of $m$ vs $t$ for $1/s=2$ (strong-noise) and
$a \approx a_c(N)$ for (top to bottom) $N=50$, $N=500$, and
$N=50000$.}
\label{m_SCvst}
\end{figure}

 As in the other case, it is interesting to look also at the temporal
 evolution of $m$. In the strong-noise regime, since the exponent
 $\theta$ is different in the NF solution and in the zero-dimensional
 regime, the crossover between the two regimes results in a crossover
 in the decay of $m(t)$.  This is evident from Fig.~\ref{m_SCvst}. The
 effective exponent switches from a value close to the prediction
 $\theta=1/s=2$ for short times, to the zero-dimensional value
 $\theta=1/2$, for longer times.

\subsection{Discussion}

  Let us underline the non-commutativity of the limits $a \rightarrow
  a_c$ and $N\rightarrow \infty$. It is only when the thermodynamic
  limit is taken first and a homogeneous mean-field $M$ is
  considered, that the NF solution is recovered. On the other hand, if
  one takes first the limit $a \rightarrow a_c(N)$ for a generic finite
  value of $N$, the zero-dimensional solution always dominates the
  scaling nearby the critical point, no matters how large
  $N$. Therefore, the inclusion of fluctuations has a dramatic effect
  on the NF solution.

  Such a conclusion holds for all the regimes considered: weak,
  intermediate, and strong-noise. This is due to the fact that the two
  aforementioned limits do not commute, implying the presence of a
  non-analyticity of the most general solution in a neighborhood of
  the critical point. This is analogous to the observation of Gaussian
  scaling in standard phase transitions whenever the system size is
  not infinite; it is only when the thermodynamic limit is taken that
  the true asymptotic scaling emerges. In the case studied here the
  role of the Gaussian scaling is replaced by a MN zero-dimensional
  non-trivial scaling.  This breaking down of the thermodynamic-limit
  behavior for finite systems has, in principle, nothing to do with
  the previously constructed Ginzburg criterion which leads to a
  breakdown of the mean-field solution only at the strong-noise
  regime.
    
  In the appendix we present a general scaling theory accounting in a
  compact form for all the previously discussed phenomenology.

\section{Finite connectivity: Random Regular Graphs}
\label{SecRRG}

  In order to shed some light on the question of whether the NF
  solution holds or not for an arbitrarily large space dimensionality,
  $d$, in which the number of sites seen by any given one is finite
  ($2d$ for a hyper-cubic lattice) we should first answer the
  following question: does the NF behavior emerge because the size of
  the system goes to infinity or because the number of nearest
  neighbors diverges?  (Note that in the fully connected graph these
  two limits coincide). In order to clarify this point we have
  considered the MN on a connected Regular Random Graph (RRG), where
  each site has fixed degree $k > 2$ and random connections (for $k=2$
  we have a one-dimensional lattice). In this case, as the number of
  neighbors is finite for each site, we expect fluctuations in $M$ and
  therefore a possible breakdown of the NF solution (at least in the
  strong-noise limit), even if the large system size limit is taken.

\begin{figure}
 \includegraphics[angle=0,width=8cm,clip]{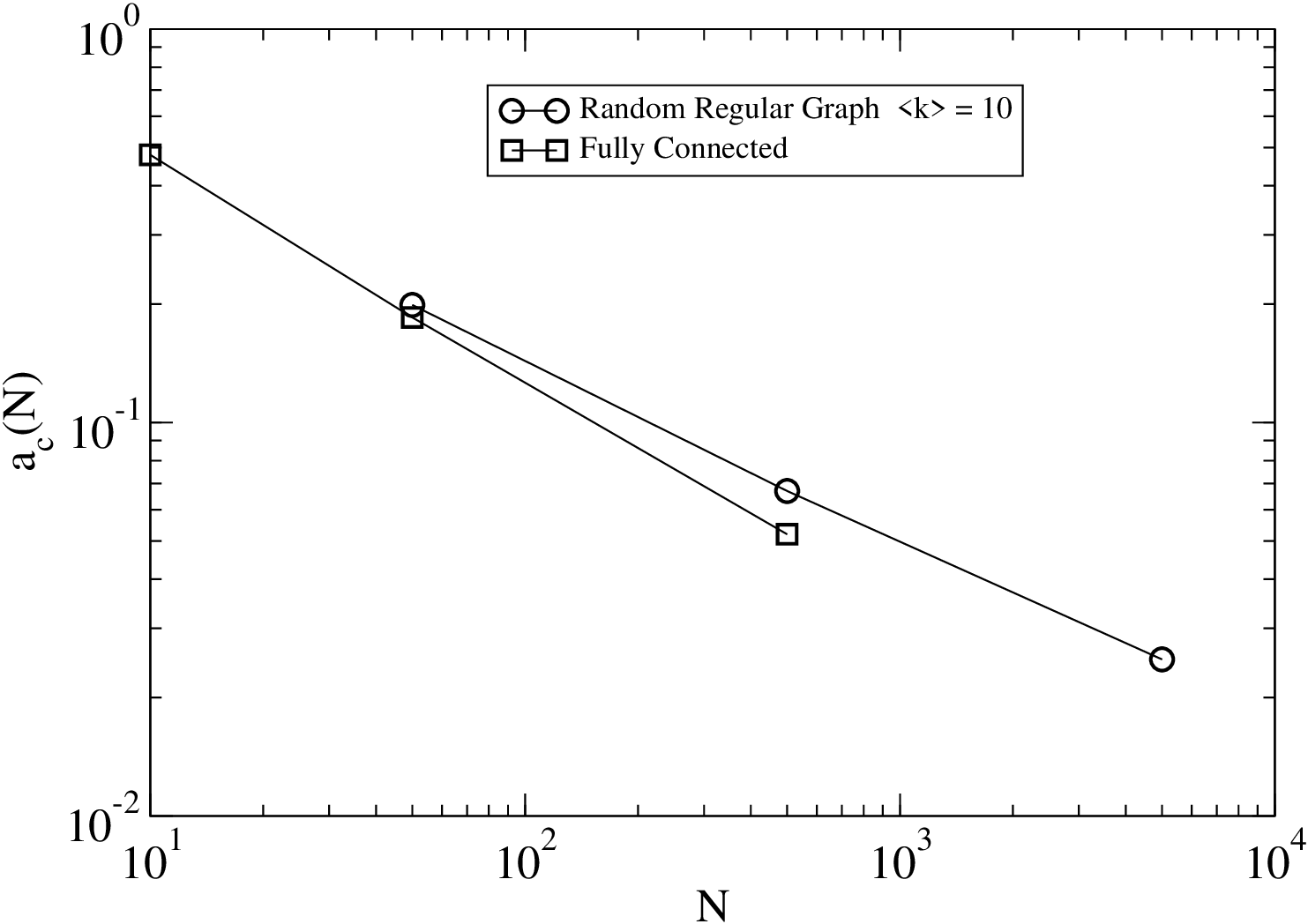} 
\caption{Plot
 of $a_c(N)$ versus $N$ for $1/s=2$ (strong-noise) for the Fully
 Connected and the RRG.}  \label{RRGa_cvsN} \end{figure}

  However, numerical results disprove such a conjecture, as we show in
  what follows. We have performed simulations of a system with $ k
  =10$ and growing $N$. It turns out that the position of the critical
  point $a_c(N)$ depends on $N$ and does not reach a finite value
  dependent only on $k$ (see Fig.~\ref{RRGa_cvsN}). Its behavior is
  not very different from what occurs on the fully connected system
  (see Eq.~\ref{a_cSC}).
 
\begin{figure}
 \includegraphics[angle=0,width=8cm,clip]{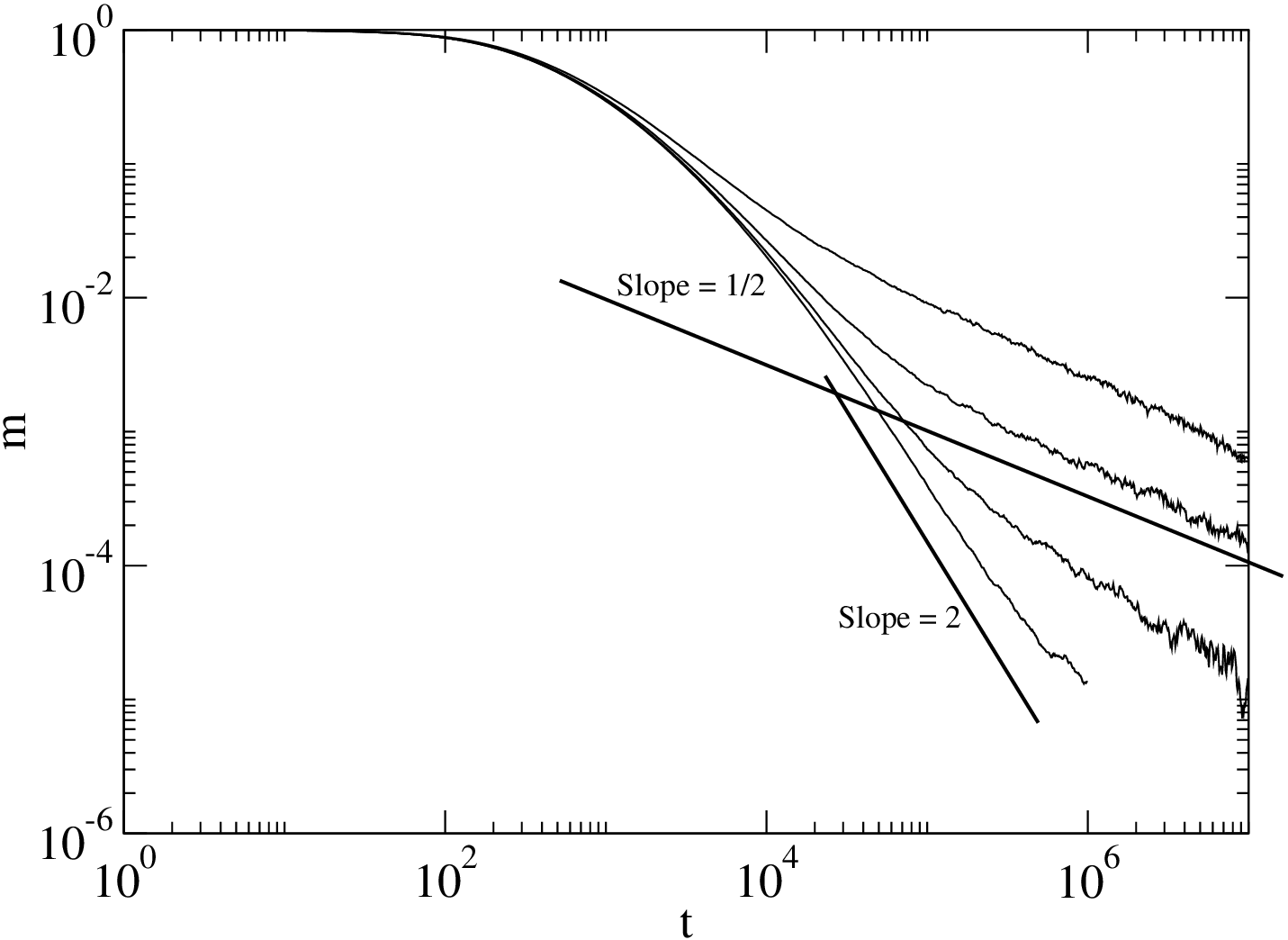}
\caption{Plot of $m$ vs $t$ for $1/s=2$ (strong-noise) and
$a \approx a_c(N)$ for (top to bottom) $N=50$, $N=500$, $N=5000$ and
$N=50000$ on the RRG.}
\label{RRGm_SCvst}
\end{figure}

  Moreover, one can monitor the temporal behavior of $m(t)$ for $a
  \approx a_c(N)$ (Fig.~\ref{RRGm_SCvst}). One finds a crossover from
  a NF behavior at short time to a zero-dimensional behavior at longer
  time, exactly as for the fully connected graph.

  We conclude that the thermodynamical limit is described, also for a
  regular random graph with finite connectivity, by the NF behavior.
  The zero-dimensional behavior holds only as long as criticality is
  studied for finite system sizes. A discussion of these facts is
  presented in the next section.

\section{Summary and Conclusions}

In this paper we have investigated the properties of the mean-field
solution for systems with multiplicative noise. First, we have
revisited the self-consistent solution obtained in Ref.~\cite{Birner}
by assuming that $M$ is a homogeneous non-fluctuating field.
We have shown that three regimes can actually be identified depending
on the noise amplitude $1/s=\sigma^2/(2D)$.  In the weak-noise case
($1/s < 1/p$) all moments of the order parameter distribution obey
ordinary Gaussian scaling.  In the intermediate regime, ($1/p < 1/s <
1$), Gaussian scaling holds only up to a certain order, with
multi-scaling emerging for higher moments.  In the strong-noise regime
($1/s > 1$) all moments scale with exponent $1/s+1$ except for the
first one that goes as $\epsilon^{1/s}$.

Complete absence of fluctuations for the spatial average $M$ holds
only in the thermodynamic limit, with a strictly infinite number of
nearest neighbors. In this paper we have relaxed the condition that
$M$ can assume only one (self-consistently fixed) value and let it
fluctuate: this is equivalent to considering a fully connected graph
with a finite number $N$ of nodes.

When $N$ is finite $M$ fluctuates and the critical behavior is not
described anymore by the NF thermodynamic limit, in none of the three
previously described regimes. While in the NF case, criticality is
governed by fluctuations of the $\phi$ field in a single realization
around its mean value, for finite $N$ what actually matters are the
fluctuations of the self-consistent field $M$. They are described by
an effective equation for a zero-dimensional system with
multiplicative noise, which exhibits a critical behavior distinct from
the one displayed in the thermodynamic limit.

The crossover between the two types of behavior is well described by
the zero-dimensional description: both the crossover point and the
critical point location depend on $N$ so that for any finite value of
$N$ there is a finite interval of values of the control parameter
where the effective zero-dimensional behavior is observed.

Finally, by means of numerical simulations on connected Random Regular
Graphs, we have also shown that all these results still hold (at least
qualitatively) for random regular graphs with finite connectivity.
This is somehow counterintuitive, and seems to contradict the
previously introduced Ginzburg criterion for the strong-noise
regime. Indeed, given that the fluctuations of $M$ are predicted by
the Ginzburg criterion to be relevant in the strong-noise regime of
the NF solution when the number of nearest neighbors is not infinity,
it is hard to understand, why the solution in the random regular
graph, in which $M_i$ is the fluctuating average taken over the $k$
neighbors of any given site $i$, behaves so similarly (for any value
of $N$) to the solution in the fully connected network in which
$M=M_i$ is fluctuation-less. 

We do not have a clear explanation of this fact, but we believe that
this is so because of the small-world \cite{SW} nature of the RRG
topology. The small-world property implies that one can reach any
arbitrary node starting from a generic site with a small number of
steps following network links. In this way, every site is nearby any
other one, making it difficult to create ``local patches'' with an
over-density or sub-density, which would give rise to inhomogeneities
and a broad field distribution. At this point, it would be interesting
to study the MN equation on a Cayley tree to see if, by introducing
well defined spatial neighborhoods, the previous results and
interpretation are sustained.  We are presently analyzing such a
problem, but prefer to leave the delicate issues involved in such a
study for a future publication.

Many interesting questions remain to be answered. The main one is
whether the obtained NF solution in the strong-noise regime is valid
in arbitrarily large but finite physical dimensions, or whether it
emerges only in $d=\infty$ (which is obviously related to the
existence of a finite upper critical dimension for MN and KPZ
problems). The Ginzburg criterion for the strong-noise solution seems
to point out to the second possibility, while the fact that in the
finite-connectivity random regular graphs the NF solution emerges in
the thermodynamic limit, could be interpreted as supporting the first
one. Further analysis along these lines is left for future work.

It remains also to be understood which is the fate of the
intermediate-noise regime once the MN equation is embedded into a
$d$-dimensional lattice. Does it simply disappear? Does it survive,
introducing some type of anomalous effect for high-order moments?
Does it have any analogous in KPZ-like systems?

It is our hope that this work will stimulate future research in this
exciting and ever surprising field of systems with multiplicative
noise.

\vspace{0.25cm}

\section*{Acknowledgments}{\nonumber}

We acknowledge useful discussions with L. Pietronero, and
P. L. Garrido. We specially thank A. Gabrielli who participated in the
early stages of this project, and G. Parisi for useful comments on the
first version of the manuscript. M.~A.~M. acknowledges financial
support from the Spanish MCyT (FEDER) under project BFM2001-2841, and
from the Acci\'on Integrada Hispano-Italiana HI2003-0344.

\vspace{0.20cm}

\section*{Appendix: General scaling of the distributions}{\nonumber}

It is possible to summarize all the scaling regimes described in the
paper in a general scaling form of the distributions.  This describes
the crossover between the zero-dimensional case and the thermodynamic
limit as $N$ grows.  The equations obeyed by the distribution
$R(\phi|M;a,N)$ of the $\phi$ field conditional to a value $M$ and by
the full distribution $P(\phi;a,N)$ are
\be
\left \{\begin{array}{lcl}
R(\phi|M;a)& \propto & \phi^{-\alpha}
\exp\left(- {b s \phi^p \over p D} - {M s \over \phi } \right) \\
P(\phi;a,N)  & = & \int_0^{\infty} dM 
R(\phi|M;a) Q(M;a,N).
\end{array}
\right.
\ee
The solution of these equations found assuming no fluctuations for $M$ is
\be
\left \{\begin{array}{lcl}
R(\phi;M)& \propto & \phi^{-\alpha}
\exp\left(- {b s \phi^p \over p D} - {M s \over \phi } \right) \\
Q(M;a,N)&\approx & \delta[M-m(a)] \\
P(\phi;a,N)  & = & R(\phi|M)|_{M=m(a)} 
\end{array}
\right.
\ee
A more general explicit ansatz for the solution to this set of
equations, valid for generic $N$, is
\be
\left \{\begin{array}{lcl}
R(\phi|M)& \propto &  M^{\alpha-1} \phi^{-\alpha}
\exp\left(- {b s \phi^p \over p D} - {M s \over \phi } \right)\\
Q(M;a,N)&\approx & M^{-\alpha'} e^{-\left({M \over M_u}\right)^2} \\
P(\phi;a,N)  & \sim & \left \{ \displaystyle 
\begin{array}{ll}
{\phi^{-\alpha'} \over \alpha-\alpha'} & \phi < M_u(a)\\
{\phi^{-\alpha} e^{-b s \phi^p \over p D} \over \alpha-\alpha'} 
M_u^{\alpha-\alpha'}& \phi > M_u(a)
\end{array}
\right.
\end{array}
\right.  \ee 
where $\alpha'=2 [1-a/\sigma^2(N)]$.  The form of $Q(M)$ encodes the
different critical behaviors.  In the NF limit ($N$ going to $\infty$
first) $\alpha'$ is negative. In this way $Q(M)$ tends to
$\delta(M-M_u)$; moments go to zero because $M_u$ goes to zero as
$a^\beta$.  In the opposite nontrivial limit ($\epsilon$ going to zero
with $N$ fixed) $\alpha'$ is positive and grows up to $1$ for $a \to
a_c(N)$.  In this case, the moments go to zero because the
distribution $Q(M)$ develops a peak in zero.  The critical behavior of
the moments in this limit is governed by the way $\alpha'$ goes to
$1$. The origin of the two limits ($\epsilon \rightarrow 0$ and $N
\rightarrow \infty$) non-commutativity has its roots in the
non-analytical form of $Q(M$ at $M=0$.


\begin{thebibliography}{99}
\bibitem{footnote}  Though the validity of naive power-counting is
not clear for the MN equation, this scaling relation seems to be
preserved in all the cases we will discuss in what follows.
\bibitem{GGC}
Note that this is slightly different from the standard form of the
Ginzburg criterion \cite{Birgenau,LB}, in which one typically
determines the spatial dimension at which a Gaussian solution breaks
down owing to the introduction of spatial fluctuations. To construct a
criterion along such a line, one needs to have a mean-field estimation
of the exponents $\gamma$ and $\nu$ associated with the susceptibility
and the correlation length respectively. In the absence of a general
mean-field theory to compute them we leave this as an open task.
\end{thebibliography}
\end{document}